# *Geometry and Topological photonics*

*Mário G. Silveirinha*[*]

[(1)] *University of Lisbon–Instituto Superior Técnico and Instituto de Telecomunicações, Avenida Rovisco Pais, 1, 1049-001 Lisboa, Portugal,* mario.silveirinha@co.it.pt

**Abstract**

Topological photonics provides a powerful framework to describe and understand many nontrivial wave phenomena in complex electromagnetic platforms. The topological index of a physical system is an abstract global property that depends on the family of operators that describes the propagation of Bloch waves. Here, we highlight that there is a profound geometrical connection between topological physics and the topological theory of mathematical surfaces. We show that topological band theory can be understood as a generalization of the topological theory of surfaces and that the genus of a surface can be regarded as a Chern number of a suitable operator defined over the surface. We point out some nontrivial implications of topology in the context of radiation problems and discuss why for physical problems the topological index is often associated with a bulk-edge correspondence.

---

[*] To whom correspondence should be addressed: E-mail: mario.silveirinha@co.it.pt

## I. Doughnuts and coffee cups

Topology studies the properties of objects that remain invariant under a smooth transformation. When some property of a mathematical object is unaffected by a deformation it is called a topological invariant. The concept is particularly simple to visualize in the case of geometrical surfaces. In such a case, the number of "holes" in the surface is a topological invariant as any smooth deformation of the surface preserves the number of holes, which is thereby a global property fully independent of any local features of the surface. In the theory of surfaces, the topological invariant is known as the genus (see Fig. 1). In particular, one can say that a "doughnut" is topologically equivalent to a coffee cup with one handle, but it is topologically distinct from a soccer ball.

Topological ideas are of central importance in many physical systems. For example, in the low temperature limit, the conductivity of magnetically-biased materials with completely filled electronic bands is determined by a topological invariant. This property is the basis of the integer quantum Hall effect [1, 3]. The quantized Hall conductivity is profoundly related to the emergence of electronic surface states in the material. A similar effect is observed in generic wave platforms (either fermionic or bosonic) invariant under discrete translations of space (e.g., photonic crystals, phononic crystals, etc) [4-14], and sometimes in platforms with a continuous translation symmetry [15-17]. In fact, in physical systems there is a bulk-edge correspondence that relates the bulk topological invariant with the propagation of edge-states at the boundary [18-21].

For Chern insulators, the topological invariant can be expressed in terms of some abstract quantity known as the Berry connection or Berry curvature [4, 5, 6]. It is well known, and widely discussed in many review articles, that there is a close parallelism between topological band theory and the topological theory of surfaces [6]. It is however less well known that topological band theory can be regarded as a generalization of the topological theory of surfaces (Fig. 1). The key objective of this article is to fill in such a gap in the photonics literature, and present an explicit elementary and pedagogical connection between the two frameworks, highlighting that the topological theory of surfaces corresponds to a particular case of topological band theory. In addition, we discuss why for physical systems the topological index is typically associated with a bulk edge correspondence, and moreover, we point out some nontrivial implications of topology in the context of the polarization of a generic antenna.

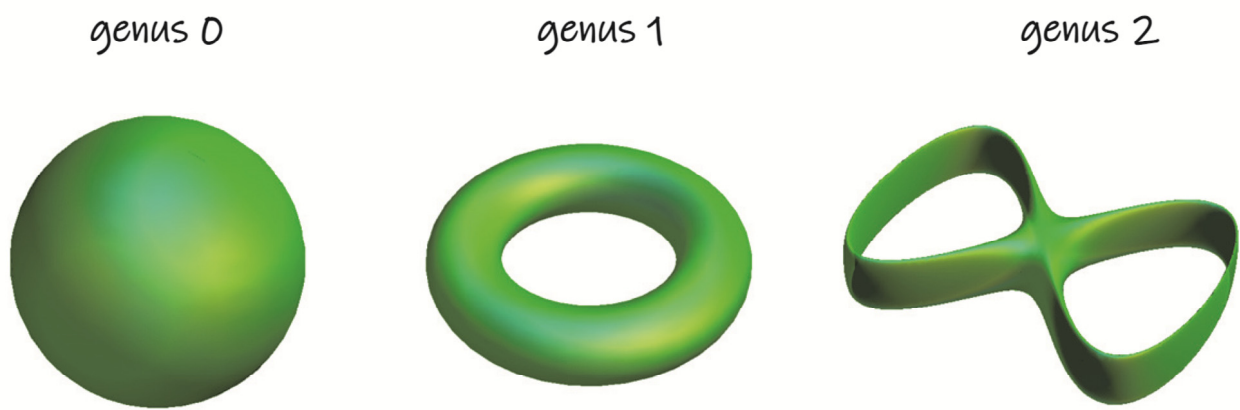

**Fig. 1** Examples of closed surfaces associated with different topologies. A smooth deformation of any of the surfaces preserves the number of holes (genus), which is thereby a topological invariant.

## II. Topological theory of surfaces

In this section, we present an elementary and pedagogical introduction to the topological theory of mathematical surfaces with the intent to point out its profound connection with topological photonics.

### A. *Gauss-Bonnet theorem*

The genus of a surface embedded in a 3D space is determined by the number of holes in the surface (Fig. 1). The genus is a very intuitive property as it can be geometrically visualized. Notably, the genus $g$ can be as well expressed in terms of the local "curvature" of the surface $S$. The Gauss-Bonnet theorem establishes that for a closed and connected surface [22, 23]:

$$\frac{1}{4\pi}\int_S \mathcal{K}ds = (1-g) \qquad (1)$$
$$= 1, 0, -1, -2, ...$$

Here, $\mathcal{K}$ is the so-called Gaussian curvature of the surface which is related to the principal curvature radii $R_1, R_2$ at a given point as $\mathcal{K}=1/(R_1 R_2)$ [22, 24]. For a spherical surface the curvature radii coincide with the radius of the sphere. The Gaussian curvature is positive for convex surfaces and negative for concave surfaces. It vanishes for planar surfaces.

The Gauss-Bonnet theorem links the genus with the Gaussian curvature integrated over the entire surface. In particular, it implies that the total curvature is precisely *quantized* and must be an integer number in units of $4\pi$. For example, a spherical surface with radius $R$ has curvature $\mathcal{K}=1/R^2$, and thereby $\int_S \mathcal{K}ds = 4\pi$ consistent with $g=0$. A smooth perturbation of a closed surface can drastically change the local Gaussian curvature, but rather remarkably the total curvature remains invariant. In the following subsections, we provide a reasonably detailed explanation why the integral of the Gaussian curvature is quantized and why it determines the genus of a surface. The results are especially relevant because they can be readily extended to topological band theory, as shown later in Sect. III.

### *B. Curvature*

It is useful to briefly review the concept of Gaussian curvature of a surface. First, consider a generic planar curve $\gamma(u)$. Let $\hat{\mathbf{n}}$ be the normal unit vector in the curve plane. The radius of curvature $R$ is defined by the geometrical construction of Fig. 2a. In order to obtain an explicit formula for the curvature $\mathcal{K}=1/R$, consider an infinitesimal displacement $dl$ along the curve. The corresponding angle $d\theta$ can be found noting that $d\theta \approx \sin(d\theta) = |\hat{\mathbf{n}}(u_0 + du) \times \hat{\mathbf{n}}(u_0)|$. This relation implies that $d\theta = \left|\dfrac{d\hat{\mathbf{n}}}{du}\right| du$. Note that $\dfrac{d\hat{\mathbf{n}}}{du}$ and $\hat{\mathbf{n}}$ are perpendicular vectors. Since the curvature radius satisfies $R d\theta = dl$, it follows that

$$\mathcal{K} dl = \pm \left|\frac{d\hat{\mathbf{n}}}{du}\right| du = \frac{d\hat{\mathbf{n}}}{du} \cdot \hat{\mathbf{t}}\, du, \qquad (2)$$

where $\hat{\mathbf{t}} = \dfrac{\boldsymbol{\gamma}'}{|\boldsymbol{\gamma}'|}$ is the unit vector tangent to the curve. The leading sign in the first identity depends if the curve is convex or concave. Note that $\dfrac{d\hat{\mathbf{n}}}{du}$ and $\hat{\mathbf{t}} = \dfrac{\boldsymbol{\gamma}'}{|\boldsymbol{\gamma}'|}$ are parallel. Clearly, the curvature is determined by derivative of $\hat{\mathbf{n}}$ along the curve. Noting that $dl = |\boldsymbol{\gamma}'| du$, one obtains the following explicit formula for the curvature: $\mathcal{K} = \dfrac{(d\hat{\mathbf{n}}/du) \cdot \boldsymbol{\gamma}'}{\boldsymbol{\gamma}' \cdot \boldsymbol{\gamma}'}$.

Let us now consider a closed surface $S$ embedded in three-dimensional space. The surface is parameterized by the mapping $\mathbf{r} = \mathbf{r}(u_1, u_2)$. The parameters $u_1, u_2$ give the coordinates of a generic surface point. The outward oriented normal unit vector is denoted by $\hat{\mathbf{n}}$ (Fig. 2b). The vectors $\dfrac{\partial \mathbf{r}}{\partial u_1}$ and $\dfrac{\partial \mathbf{r}}{\partial u_2}$ are tangent to the surface and generate the tangent plane at the point $\mathbf{r}(u_1, u_2)$. For convenience, we assume an orthogonal system of coordinates such that $\dfrac{\partial \mathbf{r}}{\partial u_1}$ and $\dfrac{\partial \mathbf{r}}{\partial u_2}$ are perpendicular at every point of the surface. Furthermore, we introduce the unit tangent vectors $\hat{\mathbf{t}}_i = \partial_i \mathbf{r} / h_i$ ($i=1,2$) with $h_i = |\partial_i \mathbf{r}|$ and $\partial_i \equiv \partial / \partial u_i$. Without loss of generality, it is supposed that $\hat{\mathbf{n}} = \hat{\mathbf{t}}_1 \times \hat{\mathbf{t}}_2$.

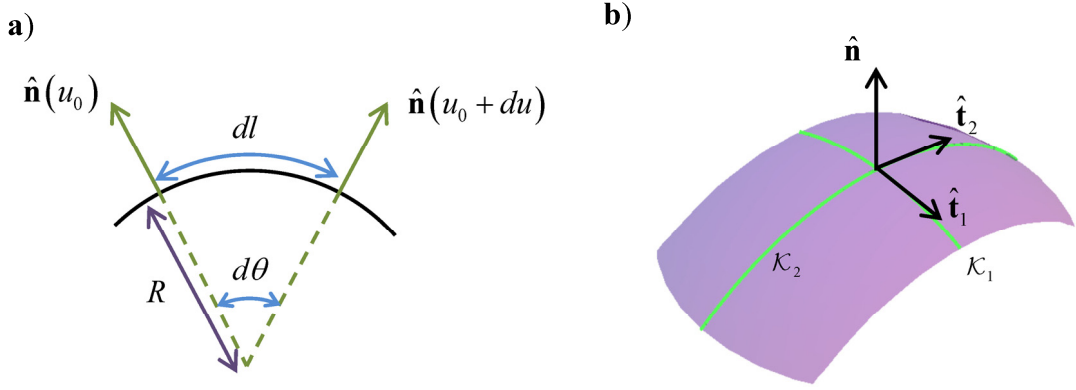

**Fig. 2 a)** The radius of curvature of a smooth planar curve is determined by the derivative of the normal unit vector $\hat{\mathbf{n}}$. **b)** Principal curvatures and principal directions (in green) of a generic surface. In the figure, $\hat{\mathbf{t}}_1$ and $\hat{\mathbf{t}}_2$ are supposed to be aligned with the principal directions.

Since $\hat{\mathbf{n}} \cdot \hat{\mathbf{n}} = 1$, the derivative of the normal vector along $u_i$, $\partial \hat{\mathbf{n}} / \partial u_i$ ($i$=1,2), is a vector of the tangent space: $\dfrac{\partial \hat{\mathbf{n}}}{\partial u_i} \cdot \hat{\mathbf{n}} = 0$. Thus, it is always possible to write:

$$\begin{pmatrix} \dfrac{\partial \hat{\mathbf{n}}}{\partial u_1} \\ \dfrac{\partial \hat{\mathbf{n}}}{\partial u_2} \end{pmatrix} = \underbrace{\begin{pmatrix} N_{11} & N_{12} \\ N_{21} & N_{22} \end{pmatrix}}_{\mathbf{N}} \begin{pmatrix} \hat{\mathbf{t}}_1 \\ \hat{\mathbf{t}}_2 \end{pmatrix}, \qquad N_{ij} = \dfrac{\partial \hat{\mathbf{n}}}{\partial u_i} \cdot \hat{\mathbf{t}}_j. \qquad (3)$$

Consider now some curve $\gamma(u)$ contained in the surface $S$ and passing through the point of interest. Different from the case of Fig. 2a, in general the derivative of $\hat{\mathbf{n}}$ along the curve ($\dfrac{d\hat{\mathbf{n}}}{du}$) does not need to be parallel to the corresponding tangent vector $\hat{\mathbf{t}} = \dfrac{\gamma'}{|\gamma'|}$. A detailed analysis shows that there are precisely two *principal directions* in the tangent plane along which $\dfrac{d\hat{\mathbf{n}}}{du}$ and $\hat{\mathbf{t}}$ are collinear. The principal directions are always orthogonal and determine the two principal curvatures $\mathcal{K}_1, \mathcal{K}_2$.

Let us first suppose that the two principal directions at the point of interest are aligned with $\hat{\mathbf{t}}_1, \hat{\mathbf{t}}_2$, as illustrated in Fig. 2b. In that situation, the matrix $\mathbf{N}$ must be diagonal. Then, in analogy with Eq. (2), one can write $\mathcal{K}_i dl_i = \frac{\partial \hat{\mathbf{n}}}{\partial u_i} \cdot \hat{\mathbf{t}}_i du_i = N_{ii} du_i$ ($i$=1,2). The Gaussian curvature is by definition the product of the two principal curvatures $\mathcal{K} = \mathcal{K}_1 \mathcal{K}_2$, so that:

$$\mathcal{K} ds = \det \mathbf{N} \, du_1 du_2. \tag{4}$$

In the above, $ds = dl_1 dl_2$ is the element of area in the surface and it was taken into account that $N_{11} N_{22}$ coincides with the determinant of the matrix when $\mathbf{N}$ is diagonal.

In the general case, the problem of finding the principal directions can be reduced to an eigenvalue problem (details are omitted for conciseness) [22]. The Gaussian curvature is still determined by Eq. (4) in the general case. Noting that $ds = h_1 h_2 du_1 du_2$, the Gaussian curvature can be explicitly written as $\mathcal{K} = \det \mathbf{N} / (h_1 h_2)$.

## C. The Gaussian curvature as the derivative of a potential

Interestingly, the Gaussian curvature can be expressed as the derivative of some function. Indeed, using Eq. (3) and noting that $N_{ij} = -\hat{\mathbf{n}} \cdot \frac{\partial \hat{\mathbf{t}}_j}{\partial u_i}$ one can write $\det \mathbf{N} = \langle \partial_1 \hat{\mathbf{t}}_1 | \hat{\mathbf{n}} \rangle \langle \hat{\mathbf{n}} | \partial_2 \hat{\mathbf{t}}_2 \rangle - \langle \partial_1 \hat{\mathbf{t}}_2 | \hat{\mathbf{n}} \rangle \langle \hat{\mathbf{n}} | \partial_2 \hat{\mathbf{t}}_1 \rangle$. The formula $N_{ij} = -\hat{\mathbf{n}} \cdot \frac{\partial \hat{\mathbf{t}}_j}{\partial u_i}$ can be derived by differentiating the equation $\hat{\mathbf{n}} \cdot \hat{\mathbf{t}}_j = 0$ with respect to $u_i$. For convenience, we introduced the canonical inner product of two vectors defined as $\langle \mathbf{v} | \mathbf{w} \rangle = \mathbf{v}^* \cdot \mathbf{w}$. Next, we note that the normal and tangent vectors satisfy the completeness relation $\mathbf{1} = |\hat{\mathbf{n}}\rangle\langle\hat{\mathbf{n}}| + |\hat{\mathbf{t}}_1\rangle\langle\hat{\mathbf{t}}_1| + |\hat{\mathbf{t}}_2\rangle\langle\hat{\mathbf{t}}_2|$. Hence, using $|\hat{\mathbf{n}}\rangle\langle\hat{\mathbf{n}}| = \mathbf{1} - |\hat{\mathbf{t}}_1\rangle\langle\hat{\mathbf{t}}_1| - |\hat{\mathbf{t}}_2\rangle\langle\hat{\mathbf{t}}_2|$ and taking into account that $\langle \partial_i \hat{\mathbf{t}}_j | \hat{\mathbf{t}}_j \rangle = 0$ ($i,j$=1,2) one finds that:

$$\begin{aligned}\det \mathbf{N} &= \langle \partial_1 \hat{\mathbf{t}}_1 | \partial_2 \hat{\mathbf{t}}_2 \rangle - \langle \partial_1 \hat{\mathbf{t}}_2 | \partial_2 \hat{\mathbf{t}}_1 \rangle \\ &= \partial_1 A_2 - \partial_2 A_1 \end{aligned} \tag{5}$$

where $(A_1, A_2)$ are defined in terms of the complex vector field $\mathbf{f} = \frac{1}{\sqrt{2}}(\hat{\mathbf{t}}_1 - i\hat{\mathbf{t}}_2)$ as follows:

$$A_j = \langle \mathbf{f} | i\partial_j \mathbf{f} \rangle, \quad j=1,2. \tag{6}$$

This demonstrates that $\det \mathbf{N}$ is the derivative of some function. It is underlined that the derivation assumes an orthogonal system of coordinates. The complex vector field $\mathbf{f}$ is normalized as $\langle \mathbf{f} | \mathbf{f} \rangle = 1$. Using this property, it is straightforward to show that $A_j$ is a real-valued function.

The Gaussian curvature $\mathcal{K} = \det \mathbf{N}/(h_1 h_2)$ can be written in terms of the surface divergence (Div) of a tangent vector field $\mathcal{A}$ as

$$\mathcal{K} = \mathrm{Div}(\mathcal{A} \times \hat{\mathbf{n}}), \tag{7}$$

where $\hat{\mathbf{n}}$ is the outward unit normal vector. The vector $\mathcal{A}$ depends on $\hat{\mathbf{t}}_1$ and $\hat{\mathbf{t}}_2$, which are unit vectors tangent to the surface determined by the system of orthogonal coordinates. It is explicitly given by $\mathcal{A} = \langle \mathbf{f} | i\,\mathrm{Grad}\,\mathbf{f} \rangle \equiv \hat{\mathbf{t}}_1 \frac{1}{h_1} \langle \mathbf{f} | i\,\partial_1 \mathbf{f} \rangle + \hat{\mathbf{t}}_2 \frac{1}{h_2} \langle \mathbf{f} | i\,\partial_2 \mathbf{f} \rangle$ or equivalently $\mathcal{A} = \frac{A_1}{h_1}\hat{\mathbf{t}}_1 + \frac{A_2}{h_2}\hat{\mathbf{t}}_2$. Here, Div and Grad stand respectively for the surface divergence and gradient operators [25]. Note that for a generic tangent vector field $\mathcal{B} = \mathcal{B}_1 \hat{\mathbf{t}}_1 + \mathcal{B}_2 \hat{\mathbf{t}}_2$ one has $\mathrm{Div}(\mathcal{B}) = \frac{1}{h_1 h_2} \frac{\partial}{\partial u_1}(h_2 \mathcal{B}_1) + \frac{1}{h_1 h_2} \frac{\partial}{\partial u_2}(h_1 \mathcal{B}_2)$. It will be shown in section III that the vector field $\mathcal{A}$ is closely related to the Berry potential of topological band theory.

### D. Genus as an obstruction

The fundamental theorem of calculus relates the integral of a function derivative with the values of the function at the boundary. The Stokes theorem generalizes this result to generic surfaces as [25]:

$$\iint_S ds\, \text{Div}(\mathcal{A} \times \hat{\mathbf{n}}) = \oint_{\partial S} \mathcal{A} \cdot \mathbf{dl}. \tag{8}$$

Here, $\mathcal{A}$ is a generic tangent vector field, $S$ is the surface of integration and $\partial S$ is the respective boundary curve. The orientation of $\partial S$ is locked to the orientation of $\hat{\mathbf{n}}$, according to the usual right-hand rule. Thereby, the Stokes theorem implies that the total Gaussian curvature of some surface patch $M$ satisfies:

$$\int_M \mathcal{K} ds = \oint_{\partial M} \mathcal{A} \cdot \mathbf{dl}. \tag{9}$$

The total Gaussian curvature of the surface patch $M$ only depends on the behavior of the "potential" $\mathcal{A}$ at the boundary! In particular, the surface patch can be arbitrarily deformed away from the boundary region without changing the total curvature of the surface patch. When Eq. (7) holds over the entire surface, i.e., when the surface can be parameterized with a single globally defined smooth mapping, it is possible to take $M = S$. In these circumstances, because a closed surface has no boundary, the line integral $\oint_{\partial S} \mathcal{A} \cdot \mathbf{dl}$ vanishes and $\int_S \mathcal{K} ds = 0$, i.e., the total curvature is zero.

For example, a torus with large radius $R$ and small radius $r$ can be parameterized as

$$\mathbf{r}(\varphi, \phi) = \left((R + r\cos\phi)\cos\varphi, (R + r\cos\phi)\sin\varphi, r\sin\phi\right), \tag{10}$$

with $(\varphi, \phi) \in [0, 2\pi]^2$ a system of orthogonal coordinates. The tangent vectors are $\hat{\mathbf{t}}_1 = h_1^{-1} \partial \mathbf{r} / \partial \varphi$ and $\hat{\mathbf{t}}_2 = h_2^{-1} \partial \mathbf{r} / \partial \phi$ with $h_1 = |\partial \mathbf{r} / \partial \varphi|$ and $h_2 = |\partial \mathbf{r} / \partial \phi|$. Both $\hat{\mathbf{t}}_1$ and $\hat{\mathbf{t}}_2$ are globally defined on the surface of the torus as smooth functions, and thereby $\mathcal{A}$ has the same property. Hence, from the discussion of the previous paragraph, we conclude that the total curvature of the torus vanishes ($g = 1$). The trefoil knot is another example of a surface that can be parameterized with a single mapping, and which thereby has total curvature equal to zero.

Clearly, a nontrivial value of the total Gaussian curvature ($g \neq 1$) implies that it is impossible to parameterize the entire surface with a single smooth mapping. From a different perspective, one may say that a genus $g \neq 1$ determines an obstruction to the application of the Stokes theorem (9) with $M = S$.

## *E. Cutting and sewing*

Consider the eight-shaped surface in the top left of Fig. 3. Suppose that a sphere with a suitable radius is inserted into the central region that connects the two halves of the eight-shaped surface as illustrated in the middle panel of Fig. 3. Each half of the sphere can be sewed to one-half of the original surface to form two torus-shaped surfaces (bottom panel of Fig. 3). The total curvature of the eight-shaped surface plus the total curvature of the sphere ($4\pi$) is then identical to the total curvature of the two tori ($2 \times 0 = 0$). Thus, the total curvature of the original surface is $\int_S \mathcal{K} ds = -4\pi$.

The construction can be readily extended to a surface with $g$ holes. In the general case, by adding $g-1$ suitably positioned "spheres" one ends up with $g$ tori, with total curvature equal to zero. Thus, the total curvature of a surface with $g$ holes is $\int_S \mathcal{K} ds = -4\pi(g-1)$, in agreement with the Gauss-Bonnet theorem.

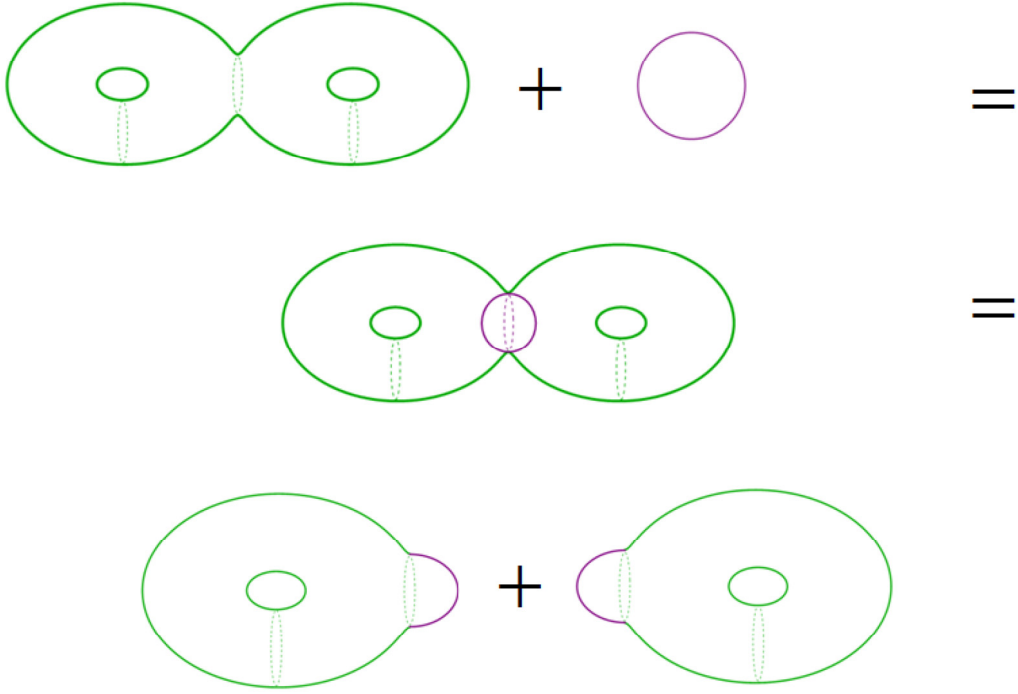

**Fig. 3** An eight-shaped surface with a sphere inserted in the central section may be seen as the juxtaposition of two torus-shaped surfaces. Thus, the total curvature of the eight-shaped surface is the sum of the curvatures of the two tori minus the curvature of a sphere.

## III. Link with Hermitian operators

In order to relate the topological theory of surfaces with topological band theory, next we consider the family of Hermitian operators defined over a surface $S$ such that

$$\hat{H}(\mathbf{r}) = i\hat{\mathbf{n}}(\mathbf{r}) \times \mathbf{1}, \qquad \mathbf{r} \in S, \qquad (11)$$

with $\hat{\mathbf{n}}$ the normal to the surface. Thus, each point of the closed surface $\mathbf{r}$ is associated with an operator $\hat{H}(\mathbf{r})$. The operator transforms a generic complex vector $\mathbf{v}$ into another complex vector given by $\hat{H} \cdot \mathbf{v} = i\hat{\mathbf{n}}(\mathbf{r}) \times \mathbf{v}$. The operator $\hat{H}(\mathbf{r})$ is Hermitian and (apart from the multiplication by $i$) determines a rotation of 90º about the normal unit vector $\hat{\mathbf{n}}$.

## A. Band structure

Let $\hat{\mathbf{t}}_1, \hat{\mathbf{t}}_2$ be vectors tangent to the surface such that $\hat{\mathbf{t}}_1, \hat{\mathbf{t}}_2, \hat{\mathbf{n}}$ form a orthogonal system with $\hat{\mathbf{t}}_1 \times \hat{\mathbf{t}}_2 = \hat{\mathbf{n}}$, similar to Sect. II. The spectrum of the operator $\hat{H}$ is determined by the eigenvectors $\hat{\mathbf{n}}$, $\frac{1}{\sqrt{2}}(\hat{\mathbf{t}}_1 + i\hat{\mathbf{t}}_2)$ and $\frac{1}{\sqrt{2}}(\hat{\mathbf{t}}_1 - i\hat{\mathbf{t}}_2)$, which are associated with the eigenvalues $\Omega = 0, +1, -1$, respectively. The eigenvalues are independent of the point on the surface and hence the associated "band structure" is flat (Fig. 4).

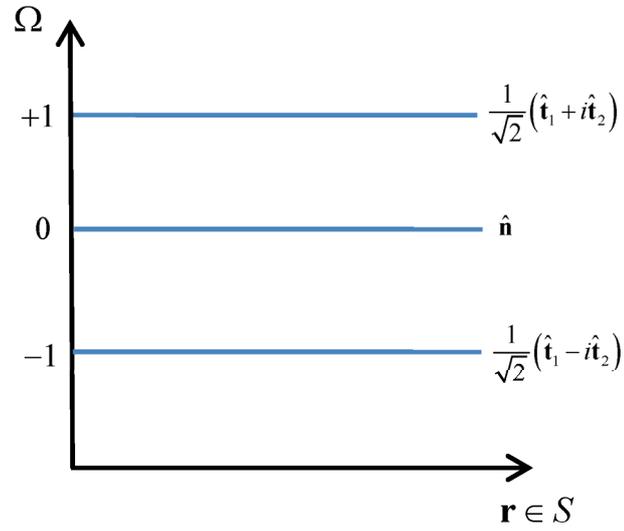

**Fig. 4** Band structure of the operator $\hat{H}(\mathbf{r}) = i\hat{\mathbf{n}}(\mathbf{r}) \times \mathbf{1}$. The spectrum of the operator is formed by 3 flat bands associated with the eigenvectors shown in the insets near each band.

## B. Berry potential, Berry curvature and gauge transformations

Let us focus on the family of eigenvectors $\mathbf{f} = \frac{1}{\sqrt{2}}(\hat{\mathbf{t}}_1 - i\hat{\mathbf{t}}_2)$ with $\Omega = -1$. Similar to section II, we introduce the vector $\mathcal{A} = \langle \mathbf{f} | i \, \mathrm{Grad}\, \mathbf{f} \rangle$, which is nothing but the standard Berry potential of topological band theory for the lowest "energy" band. Since the eigenvectors are normalized as $\langle \mathbf{f} | \mathbf{f} \rangle = 1$ the Berry potential is necessarily real-valued.

Evidently, the vectors $\hat{\mathbf{t}}_1, \hat{\mathbf{t}}_2$ are not unique: any rotation of $\hat{\mathbf{t}}_1, \hat{\mathbf{t}}_2$, e.g., $\hat{\mathbf{t}}'_1 = \cos\theta \hat{\mathbf{t}}_1 + \sin\theta \hat{\mathbf{t}}_2$ and $\hat{\mathbf{t}}'_2 = -\sin\theta \hat{\mathbf{t}}_1 + \cos\theta \hat{\mathbf{t}}_2$, yields another basis of the tangent plane. A generic rotation transforms the eigenvector $\mathbf{f}$ as $\mathbf{f} \to \mathbf{f}' = \mathbf{f}e^{i\theta}$ with $\theta = \theta(\mathbf{r})$ the rotation angle, which may depend on the surface point. A transformation of the form $\mathbf{f} \to \mathbf{f}e^{i\theta}$ is a gauge transformation. A gauge transformation preserves the normalization $\langle \mathbf{f} | \mathbf{f} \rangle = 1$. Note that the condition $\langle \mathbf{f} | \mathbf{f} \rangle = 1$ leaves an overall phase factor undetermined.

Under a gauge transformation, the Berry potential is transformed as,

$$\mathcal{A} \to \mathcal{A}' = \mathcal{A} - \text{Grad}\theta, \tag{12}$$

Thus, the Berry potential depends on the basis of tangent vectors. Because of this property, the Berry potential is gauge dependent.

If $\hat{\mathbf{t}}_1, \hat{\mathbf{t}}_2$ are the vectors obtained directly from a parameterization of the surface in the original system of orthogonal coordinates (see Sect. II), the Berry potential determines the Gaussian curvature through Eq. (7). Under a gauge transformation, $\mathbf{f} \to \mathbf{f}' = \mathbf{f}e^{i\theta}$, the right-hand side of (7) is transformed as:

$$\text{Div}(\mathcal{A}' \times \hat{\mathbf{n}}) = \text{Div}((\mathcal{A} - \text{Grad}\theta) \times \hat{\mathbf{n}}) = \text{Div}(\mathcal{A} \times \hat{\mathbf{n}}), \tag{13}$$

where we used the property $\text{Div}(\text{Grad}\theta \times \hat{\mathbf{n}}) = 0$ [25]. Therefore, $\text{Div}(\mathcal{A} \times \hat{\mathbf{n}})$ is gauge invariant. This result shows that the Gaussian curvature $\mathcal{K}$ can be calculated with an *arbitrary* basis of tangent vectors $\hat{\mathbf{t}}_1, \hat{\mathbf{t}}_2$, not necessarily associated with the original coordinate parameterization. In particular, the Gaussian curvature is fully determined by eigenfunctions with $\Omega = -1$ of the Hermitian operator $\hat{H}$.

## C. Link between the genus and the Chern number

As seen, in the previous subsection, the quantity $\mathcal{K} = \text{Div}(\mathcal{A} \times \hat{\mathbf{n}})$ is gauge invariant. It is known as the Berry curvature in topological band theory [4, 5]. For the lowest frequency band of the operator $\hat{H}(\mathbf{r}) = i\hat{\mathbf{n}}(\mathbf{r}) \times \mathbf{1}$ the Berry curvature coincides with the Gaussian curvature. The Chern index is defined by:

$$\mathcal{C} = \frac{1}{2\pi} \int_S \mathcal{K} ds. \tag{14}$$

Comparing this formula with the Eq. (1), one sees that the genus of a closed surface $S$ is related to the Chern number of the corresponding family of operators $\hat{H}(\mathbf{r}) = i\hat{\mathbf{n}}(\mathbf{r}) \times \mathbf{1}$ as:

$$\mathcal{C} = 2(1-g). \tag{15}$$

Thus, the Chern index associated with a closed surface is always an even number. The topologically trivial surface ($\mathcal{C} = 0$) has genus $g = 1$ (torus), as it corresponds to a surface with vanishing total curvature, which can be parameterized with a single mapping.

## D. The topology of a spherical surface and the hairy ball theorem

Let us apply the developed formalism to the case of a spherical surface with radius $R$. Using spherical coordinates $(\theta, \varphi, r)$, it is clear that $\hat{\mathbf{n}} = \hat{\mathbf{r}}$. The vectors $\hat{\mathbf{t}}_1, \hat{\mathbf{t}}_2$ may be chosen $\hat{\mathbf{t}}_1 = \hat{\boldsymbol{\theta}}$ and $\hat{\mathbf{t}}_2 = \hat{\boldsymbol{\varphi}}$, and then $\mathbf{f} = \frac{1}{\sqrt{2}}(\hat{\boldsymbol{\theta}} - i\hat{\boldsymbol{\varphi}})$. From here, using $\text{Grad} = \frac{1}{R}\hat{\boldsymbol{\theta}}\partial_\theta + \frac{1}{R\sin\theta}\hat{\boldsymbol{\varphi}}\partial_\varphi$, the Berry potential is found to be $\mathcal{A} = \frac{-\cot\theta}{R}\hat{\boldsymbol{\varphi}}$. The family of eigenfunctions $\mathbf{f}$ is smooth, except on the poles of the sphere, $\theta = 0, \pi$. Evidently, the Berry potential has the same property. Using (9) with $M$ the sphere surface excluding two small patches centered at the poles, it is found that the total curvature is given by the line integrals of the Berry potential along circles with infinitesimal radii centered at the poles. Each circle contributes precisely $2\pi$, so that the total curvature of the sphere is $4\pi$.

Note that a gauge transformation can make $\mathbf{f}$ smooth near the poles but will necessarily introduce a singularity at some other point on the sphere surface. Indeed, if $\mathbf{f}$ could be made smooth and globally defined, the corresponding Berry potential would have the same property, and thereby the total Gaussian curvature of the sphere would vanish, which is a false preposition. Thus, it is impossible to find some globally defined $\hat{\mathbf{t}}_1, \hat{\mathbf{t}}_2$ on the surface of a sphere with $\mathbf{f} = \frac{1}{\sqrt{2}}\left(\hat{\mathbf{t}}_1 - i\hat{\mathbf{t}}_2\right)$ a smooth function.

Furthermore, the same property implies that it is impossible to find a smooth non-vanishing tangential vector field $\hat{\mathbf{t}}_1$ defined on the surface of a sphere. Indeed, if that was possible one could define $\hat{\mathbf{t}}_2 = \hat{\mathbf{n}} \times \hat{\mathbf{t}}_1$ such that $\mathbf{f} = \frac{1}{\sqrt{2}}\left(\hat{\mathbf{t}}_1 - i\hat{\mathbf{t}}_2\right)$ would be smooth and globally defined, which as discussed above is unfeasible. This result implies that it is impossible to comb a "hairy ball" flat with all the hair strands (the analogue of a vector field) parallel to the surface (Fig. 5). This result is known as the "hairy ball theorem" or "hedgehog theorem", and was first stated by Henri Poincaré in the end of the 19[th] century [23].

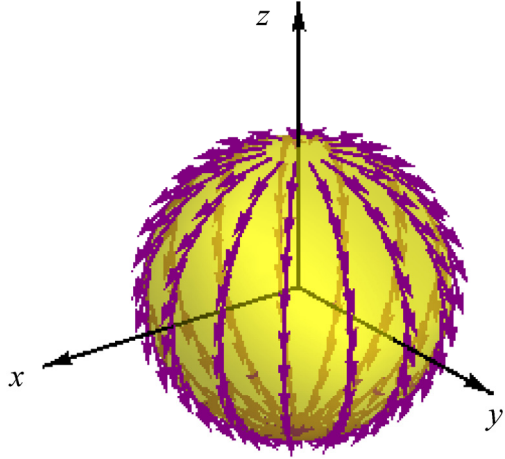

**Fig. 5** Any attempt to comb a hairy ball flat creates at least a singular point (where the vector field is not continuous) somewhere on the ball. The figure shows an attempt to comb the "hair" strands tangent to the "meridians" of the sphere, which creates singularities at both the North and South poles.

The nontrivial topology of the sphere has interesting practical consequences in the context of radiation problems. For example, consider the radiation pattern of a generic antenna under a time-harmonic excitation. For every direction of space $\hat{\mathbf{r}}$ it is possible to assign a complex vector in the tangent plane of the sphere (generated by $\hat{\boldsymbol{\theta}}, \hat{\boldsymbol{\varphi}}$) given by the electric field $\mathbf{E}$ in the far region ($\mathbf{E} \cdot \hat{\mathbf{r}} = 0$). Evidently, the electric field varies smoothly in the far-field region because it is a physical response. Let us suppose that the far-field does not have any nulls so that $|\mathbf{E}| \neq 0$. The handedness of the emitted light cannot be same for every direction of space, e.g., it is not feasible to design an antenna without nulls in the radiation pattern that emits "left" elliptical (or "left" circular) polarized light in every direction of space. In fact, if the emitted polarization is elliptical both $\text{Re}\{\mathbf{E}\}$ and $\text{Im}\{\mathbf{E}\}$ must be nonzero. If the handedness is fixed, let us say to the "left" in every direction of space, then both $\text{Re}\{\mathbf{E}\}$ and $\text{Im}\{\mathbf{E}\}$ are non-vanishing tangent vector fields over the entire surface of the "sphere". This means that $\text{Re}\{\mathbf{E}\}/|\text{Re}\{\mathbf{E}\}|$ determines a globally defined smooth mapping of the tangent space of the sphere, which as discussed previously is forbidden by the hairy ball theorem. In summary, using topological concepts it was shown that the handedness of the polarization of an antenna without any nulls in the radiation pattern cannot be fixed: there must be at least a direction of space along which the antenna polarization is linear.

An example that illustrates these ideas is the "cross-dipole" or turnstile antenna. In its simplest form it is formed by two perpendicular Hertz-dipoles fed by currents in quadrature. The radiation pattern of the cross-dipole has no nulls, but the handedness of the emitted polarization depends on the direction of observation. The plane of the dipoles divides the entire space into two hemispheres, North and South. The handedness of the emitted light (left or right) is different for the two hemispheres. In the plane of the cross-dipole, the antenna polarization is linear.

### *E. Sketch of a proof that the Chern index is an integer*

For completeness, let us present an elementary argument that illustrates why the Chern index is necessarily an integer. For simplicity we focus on the topology of the sphere (Fig. 6), but the same idea can be readily extended to other more complex geometries. As previously discussed, in general it is not possible to have a globally defined mapping of a closed surface. For example, in order to map the entire surface of a sphere one needs to consider at least two mappings, for example a mapping of the North hemisphere (top part) and a mapping of the South hemisphere (bottom part) (e.g., $\mathbf{r}(x,y) = \left(x, y, \pm\sqrt{1-x^2-y^2}\right)$).

Consider then two mappings ("top" and "bottom") of the sphere (or of a sphere-type object) and evaluate the total Gaussian curvature (Chern index) using:

$$\mathcal{C} = \frac{1}{2\pi}\int_S \mathcal{K} ds = \frac{1}{2\pi}\int_{\text{top}} \mathcal{K}_{\text{top}} ds + \frac{1}{2\pi}\int_{\text{bot}} \mathcal{K}_{\text{bot}} ds. \tag{16}$$

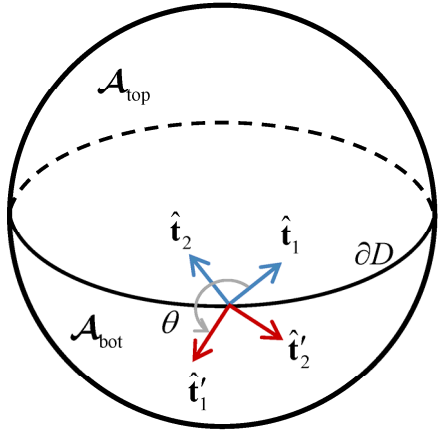

**Fig. 6** A parameterization of a spherical-type surface requires at least two mappings (e.g., of the top and bottom sections). The boundary between the bottom and top sections is denoted by $\partial D$. At the boundary $\partial D$, the two orthogonal bases of the tangent space (associated with the different parameterizations) differ by a rotation of $\theta = \theta(\mathbf{r})$. The angle $\theta = \theta(\mathbf{r})$ varies along $\partial D$.

Then, taking into account Eq. (7), $\mathcal{K}_i = \text{Div}(\mathcal{A}_i \times \hat{\mathbf{n}})$ ($i$=top, bot), it is possible to reduce the two surface integrals to line integrals over the boundary:

$$\mathcal{C} = \frac{1}{2\pi} \oint_{\partial D} \mathcal{A}_{\text{top}} - \mathcal{A}_{\text{bot}} . \qquad (17)$$

But as seen in Sect. III.B, the Berry potentials in the two sections differ by gradient of a function: $\mathcal{A}_{\text{bot}} = \mathcal{A}_{\text{top}} - \text{Grad}\,\theta$ [Eq. (12)], where $\theta = \theta(\mathbf{r})$ is the rotation angle that links the two bases of unit vectors: $\hat{\mathbf{t}}'_1 = \cos\theta\,\hat{\mathbf{t}}_1 + \sin\theta\,\hat{\mathbf{t}}_2$ and $\hat{\mathbf{t}}'_2 = -\sin\theta\,\hat{\mathbf{t}}_1 + \cos\theta\,\hat{\mathbf{t}}_2$ (the primed unit vectors are defined in the bottom section of the sphere). From this discussion, it follows that $\mathcal{C} = \frac{1}{2\pi} \oint_{\partial D} \text{Grad}\,\theta = \frac{\theta_{\text{f}} - \theta_{\text{i}}}{2\pi}$, where $\theta_{\text{f}}, \theta_{\text{i}}$ represent the rotation angles at the initial and final points of the contour $\partial D$. As the initial and final points are identical $\theta_{\text{f}}, \theta_{\text{i}}$ must differ by an integer multiple of $2\pi$. This demonstrates that the Chern index must be an integer. Note that the transformation of basis determined by the angle $\theta(\mathbf{r})$ is smooth provided the functions $\cos\theta, \sin\theta$ vary smoothly in the contour $\partial D$. Evidently, the smoothness (and continuity) of $\cos\theta, \sin\theta$ does not imply that the continuity $\theta(\mathbf{r})$. The nontrivial topology is a consequence of the discontinuity of the angle $\theta(\mathbf{r})$ (jump by a multiple of $2\pi$) in the boundary contour $\partial D$.

## IV. Topological Photonics

The geometrical concepts discussed in the previous sections can be readily generalized to the characterization of Hermitian operators. In fact, the construction of Sect. III for the operator family $\hat{H}(\mathbf{r}) = i\hat{\mathbf{n}}(\mathbf{r}) \times \mathbf{1}$ can be extended to an arbitrary (smooth) family of operators $\hat{H}(\mathbf{r})$ defined on a closed surface $S$. While the eigenfunctions of $\hat{H}(\mathbf{r}) = i\hat{\mathbf{n}}(\mathbf{r}) \times \mathbf{1}$ have a clear geometrical interpretation, for a generic $\hat{H}(\mathbf{r})$ the eigenfunctions may be abstract vectors in some abstract vector space (with an arbitrary number of dimensions).

For a generic $\hat{H}(\mathbf{r})$, the Berry potential and Berry curvature are still defined in terms of the eigenfunctions as $\mathcal{A} = \langle \mathbf{f} | i\,\mathrm{Grad}\,\mathbf{f} \rangle$ and $\mathcal{K} = \mathrm{Div}(\mathcal{A} \times \hat{\mathbf{n}})$. The Chern index is still given by Eq. (14) and remains an integer due to arguments analogous to those of Sect. III.E. For $\hat{H}(\mathbf{r}) = i\hat{\mathbf{n}}(\mathbf{r}) \times \mathbf{1}$ (specifically for the lowest frequency "band" of the operator), the Berry curvature is coincident with the Gaussian curvature, and the Chern index is determined by the genus of the surface $S$ [Eq. (15)].

In physical systems, $S$ is typically a surface in the spectral $\mathbf{k}$-space, most often a two-dimensional Brillouin zone (BZ) or in case of some Floquet topological systems a closed surface contained in a three-dimensional BZ (e.g., an isofrequency contour [26]). Thus, the family of operators is parameterized by $\mathbf{k}$: $\hat{H} = \hat{H}_\mathbf{k}$. Typically, $\hat{H}_\mathbf{k}$ is the operator that describes the propagation of the Bloch waves in a 2D periodic physical structure, for example a periodic (parallel-plate type) photonic crystal waveguide with the allowed propagation directions parallel to the *xoy* plane. Note that the two-dimensional BZ is isomorphous to a torus, i.e., to a surface with a trivial topology.

The most remarkable physical consequence of a nontrivial topology in photonics is the bulk edge correspondence [18-21]. The bulk edge correspondence links the Chern indices of the operators that describe the wave propagation in two photonic crystals with the net number of unidirectional edge states at a boundary of the two materials. The bulk edge correspondence is a consequence of the specific structure of the operators $\hat{H}_\mathbf{k}$ that determine the propagation of Bloch waves in a periodic physical platform, and has no counterpart in the topological theory of surfaces. For a photonic crystal, the operators $\hat{H}_\mathbf{k}$ are constructed from a single operator of the type $\hat{H}(-i\nabla, \mathbf{r})$ as:

$$\hat{H}_\mathbf{k} = \hat{H}(-i\nabla + \mathbf{k}, \mathbf{r}). \tag{18}$$

Due to this very specific structure, it turns out that it is possible to express the topological invariant in terms of the real-space system Green's function $\mathcal{G}$ defined as:

$$\left(\hat{H}(\mathbf{r}, -i\nabla) - \mathbf{1}\omega\right) \cdot \mathcal{G}(\mathbf{r}, \mathbf{r}', \omega) = i\mathbf{1}\delta(\mathbf{r} - \mathbf{r}'). \tag{19}$$

Specifically, the gap Chern number can be written as [21, 27, 28]:

$$\mathcal{C}_{\text{gap}} = \lim_{A_{tot} \to \infty} \frac{1}{A_{tot}} \int_{\omega_{\text{gap}} - i\infty}^{\omega_{\text{gap}} + i\infty} d\omega \iint dV dV' \left[ \text{tr}\left\{ \left[\partial_2 \hat{H} \cdot \mathcal{G}(\mathbf{r}, \mathbf{r}', \omega)\right] \cdot \left[\partial_1 \hat{H} \cdot \partial_\omega \mathcal{G}(\mathbf{r}', \mathbf{r}, \omega)\right] \right\} \right]. \tag{20}$$

Here, $A_{tot}$ is the area of a "cavity" that includes many cells of the photonic crystal, as illustrated in Fig. 7a. The gap Chern number $\mathcal{C}_{\text{gap}}$ is obtained after integration of the Green's function in the cavity (the volume integrals) and in a vertical strip of the complex plane that determines the band gap (integration in $\omega$). For further details, the reader is referred to Ref. [21].

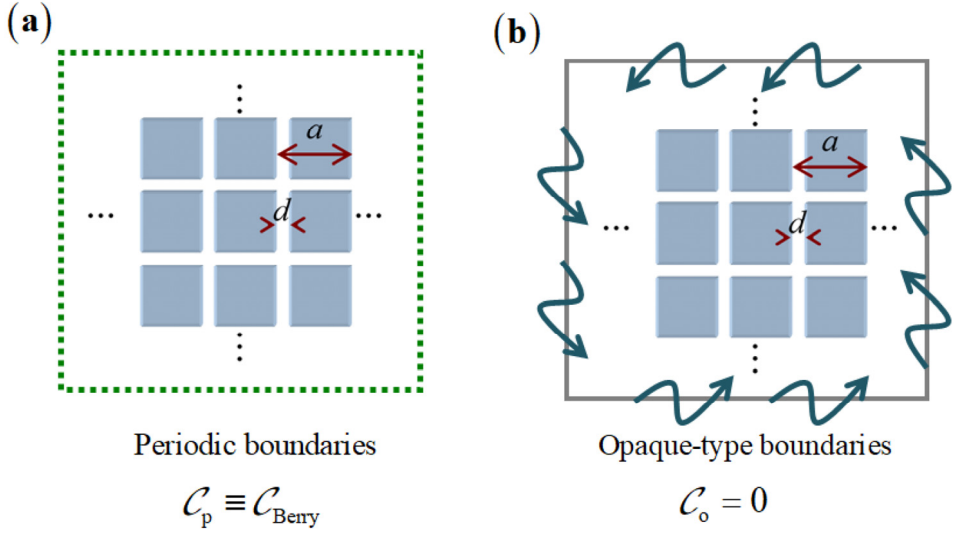

Fig. 7 The integral in Eq. (20) depends critically on the boundary conditions imposed on the cavity walls. a) For periodic boundary conditions the integral yields the usual gap Chern number. b) For opaque-type boundaries (e.g., for a perfect electric wall) it vanishes. This qualitative difference implies that in case b) the system forcibly supports edge states at the boundaries.

The Green's function in Eq. (20) must satisfy periodic boundary conditions at the cavity walls. The Green's function in real-space decays exponentially with the distance $|\mathbf{r}-\mathbf{r}'|$ source ($\mathbf{r}'$) and observation ($\mathbf{r}$) points, because in the relevant frequency range (i.e., the vertical strip of the complex frequency plane corresponding to the gap) the bulk region does not support photonic states [21]. The interesting point is that when Eq. (20) is evaluated using a Green's function that satisfies opaque-type boundary conditions (e.g., an electric wall, which is impenetrable by light) the result of the integration is exactly zero (Fig. 7b). Thus, the integral in Eq. (20) depends critically on the type of boundary conditions, even though the Green's function in the interior region of the cavity is for all purposes independent of the boundary conditions. These two features can be compatible only if the opaque-type boundaries create edge states at the cavity walls [21]. In summary, the specific structure of the operator family $\hat{H}_\mathbf{k}$ [Eq. (18)] is at the origin of the bulk-boundary correspondence.

## V. Conclusion

In this article, we revisited the link between topological band theory and the topological theory of mathematical surfaces, showing explicitly that the former can be regarded as a generalization of the latter. It particular, it was proven that the genus of a surface embedded in three-dimensional space can be related to the Chern index of a Hermitian operator family defined on the surface that rotates vectors in the tangent space by 90º. Furthermore, we discussed how the "hairy ball theorem" constrains the polarization of the wave emitted by an arbitrary antenna, showing that handedness of the polarization state of the radiated wave cannot be direction independent. Finally, we highlighted the reason why for physical systems the topological invariants are often associated with a bulk-edge correspondence, and why this property does not have a direct counterpart in the topological theory of surfaces.

**Acknowledgements:** This work is supported in part by the Institution of Engineering and Technology (IET), by the Simons Foundation, and by Fundação para a Ciência e a Tecnologia and Instituto de Telecomunicações under project UIDB/50008/2020.